\documentclass[prl,showpacs,twocolumn,superscriptaddress,preprintnumbers]{revtex4}

\usepackage{amsmath}
\usepackage{amssymb}
\usepackage{graphicx}
\usepackage{color}

\begin{document}

\title{Human behavior in Prisoner's Dilemma experiments suppresses
network reciprocity}

\author{Carlos Gracia-L\'azaro}

\affiliation{Instituto de Biocomputaci{\'o}n y F{\'\i}sica de Sistemas
Complejos (BIFI), Universidad de Zaragoza, 50018 Zaragoza, Spain}

\author{Jos\'e A.\ Cuesta}
\email{cuesta@math.uc3m.es}

\affiliation{Grupo Interdisciplinar de Sistemas Complejos (GISC),
Departamento de Matem{\'a}ticas, Universidad Carlos III de Madrid,
28911 Legan{\'e}s, Spain}

\author{Angel S\'anchez}
\homepage{http://www.anxosanchez.eu}

\affiliation{Instituto de Biocomputaci{\'o}n y F{\'\i}sica de Sistemas
Complejos (BIFI), Universidad de Zaragoza, 50018 Zaragoza, Spain}

\affiliation{Grupo Interdisciplinar de Sistemas Complejos (GISC),
Departamento de Matem{\'a}ticas, Universidad Carlos III de Madrid,
28911 Legan{\'e}s, Spain}

\author{Yamir Moreno}
\email{yamir.moreno@gmail.com}

\affiliation{Instituto de Biocomputaci{\'o}n y F{\'\i}sica de Sistemas
Complejos (BIFI), Universidad de Zaragoza, 50018 Zaragoza, Spain}

\affiliation{Departamento de F{\'\i}sica Te\'orica, Facultad de Ciencias, 
Universidad de Zaragoza, Pedro Cerbuna 12, 50009 Zaragoza, Spain}

\affiliation{Complex Networks and Systems Lagrange Lab, Institute for
   Scientific Interchange,\\ Viale S. Severo 65, 10133 Torino, Italy}

\begin{abstract}
During the last few years, much research has been devoted to strategic
interactions on complex networks. In this context, the
Prisoner's Dilemma has become a paradigmatic model,
and it has been established that imitative evolutionary dynamics lead
to very different outcomes depending on the details of the network.
We here report that when one takes into account the real behavior of
people observed in the experiments, both at the mean-field level and
on utterly different networks the observed level of cooperation
is the same. We thus show that when human subjects interact in an
heterogeneous mix including cooperators, defectors and moody conditional
cooperators, the structure of the population does not promote or inhibit
cooperation with respect to a well mixed population. 
\end{abstract}


\maketitle

In recent years, the physics of complex systems has widened its scope
by considering interacting many-particle models where the interaction
goes beyond the usual concept of force. One such line of research that
has proven particularly interesting is evolutionary game theory on
graphs \cite{szabo:2007,roca:2009a}, in which interaction between 
agents is given by a game while their own state is described by a
strategy subject to an evolutionary process
\cite{hofbauer:1998,gintis:2009}. A game that has attracted a lot of
attention in this respect is the Prisoner's Dilemma (PD) 
\cite{rapoport:1966,axelrod:1984}, a model of a situation in which 
cooperative actions lead to the best outcome in social terms, but
where free riders or non-cooperative individuals can benefit the
most individually. In mathematical terms, this is described by a
payoff matrix (entries correspond to the row player's payoffs
and C and D are respectively the cooperative and non-cooperative actions)
\begin{equation}
\begin{tabular}{c|c|c|}
\mbox{ } & C & D 
 \\ \hline
C & 1 & S
 \\ \hline
D & T & 0 \\
\hline
\end{tabular}
\label{payoffmatrix}
\end{equation}
with $T>1$ (temptation to free-ride) and $S<0$ (detriment in cooperating
when the other does not). 

In a pioneering work, Nowak and May \cite{nowak:1992} showed that the
behavior observed in a repeated Prisoner's Dilemma 
was dramatically different on a lattice than in a mean-field approach:
Indeed, on a lattice the cooperative strategy was able to prevail by
forming clusters of alike agents who outcompeted defection. 
Subsequently, the problem was considered in literally hundreds of papers
\cite{szabo:2007}, and very many differences between structured and well-mixed
(mean-field) populations were identified, although by no means they
were always in favor of cooperation \cite{hauert:2004,sysi-aho:2005}.
In fact, it has been recently realized that this problem is very sensitive
to the details of the system \cite{roca:2009a,roca:2008}, in particular
to the type of evolutionary dynamics \cite{hofbauer:2003} considered.
For this reason experimental input is needed in order to reach a sound
conclusion about what has been referred to as `network reciprocity'.

In this Letter, we show that using the outcome from the experimental 
evidence to inform theoretical models, the behavior of agents playing a
PD is the same at the mean field level and in very different networks.
To this end, instead of considering some \emph{ad hoc} imitative dynamics
\cite{nowak:1992,helbing:1992,szabo:1998}, our players will play according
to the strategy recently uncovered by Gruji\'c {\em et al.}
\cite{grujic:2010} in the largest experiment reported to date about the
repeated spatial PD, carried out on a lattice as in Nowak and May's
paper \cite{nowak:1992} with parameters  $T=1.43$ and $S=0$.

The results of the experiment were novel in several respects. First, the
population of players exhibited a rather low level of cooperation
(fraction of cooperative actions in every round of the game in the
steady state), hereafter denoted by $\langle c \rangle$. Most
important, however, was the unraveling of the structure of the strategies.
The analysis of the actions taken by the players showed a heterogeneous
population consisting of ``mostly defectors'' (defected with probability
larger than 0.8), a few ``mostly cooperators'' (cooperated with probability
larger than 0.8), and a majority of so-called moody conditional cooperators. 
This last group consisted of players that switched from cooperation to
defection with probability $P_i^{DC}=1-d-\gamma c_i=1-P_i^{CC}$ and 
from defection to cooperation with probability
$P_i^{CD}=a+\beta c_i=1-P_i^{DD}$, $c_i$ being the fraction of
cooperative actions in player $i$'s neighborhood in the previous iteration.
Conditional cooperation, i.e., the dependency of the chosen strategy on
the amount of
cooperation received, had been reported earlier in related experiments 
\cite{fischbacher:2001} and observed also for the spatial repeated PD
at a smaller scale \cite{traulsen:2010}. The new ingredient revealed in
Gruji\'c {\em et al.}'s experiment \cite{grujic:2010} 
was the dependence of the behavior on the own player's previous action,
hence the reason to call them ``moody''. Recent experiments about 
the multiplayer repeated PD confirm this observation \cite{grujic:2011b}. 

To study how the newly unveiled rules influence the emergence of
cooperation in an structured population of individuals, we first report
results from numerical simulations of a system made up of $N=10^4$
individuals who play a repeated PD game according to the experimental
observations. To this end, we explored the average level of cooperation
in four different network configurations: a well-mixed population in
which the probability that a player interacts with any other one is the
same for all players, a square lattice, an Erd\"os-Renyi (ER) graph and a
Barab\'asi-Albert (BA) scale-free (SF) network. It is worth mentioning that
the dependence on the payoff matrix only enters through the parameters
describing the players' behavior ($d$, $\gamma$, $a$, $\beta$ and the fractions
of the three types of players). Once these parameters are fixed the payoffs do
not enter anywhere in the evolution, as this is only determined by the variables
$c_i$, the local fractions of cooperative actions within each player's neighborhood.
Thus there is no possibility to explore the dependence on the payoffs because we
lack a connection between them and the behavioral parameters.

\begin{figure}
\begin{center}
\includegraphics[width=0.9\columnwidth,clip=]{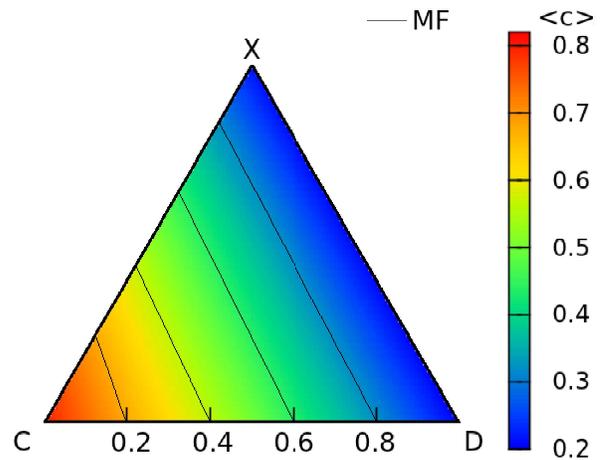}
\end{center}
\caption{(color online) Density plot of the average level of cooperation
in the stationary state,
$\langle c\rangle$, as a function of the fractions of the three strategies
(mostly cooperators, $C$, mostly defectors, $D$, and moody conditional
cooperators, $X$). The plot corresponds to a Barab\'asi-Albert network of
contacts ($\langle k\rangle=6$), but the corresponding plot for an
Erd\"os-Renyi graph or a regular lattice is indistinguishable from this
one. The system is made up of $N=10^4$ players and the rest of
parameters, taken from \cite{grujic:2010}, are: $d=0.38$, $a=0.15$,
$\gamma=0.62$, $\beta=-0.1$. The thin lines represent the mean-field
estimations [c.f.~Eq.~\eqref{SymbolicMeanFieldEquation}] for
$\langle c\rangle=0.32$, $0.44$, $0.56$, $0.68$. They very accurately match the
contour lines of the density plot corresponding to those values of
$\langle c\rangle$, thus proving that the same outcome is obtained in
a complete graph (mean-field). Simulation results have been
averaged over 200 realizations.}
\label{figure1}
\end{figure}

In Fig.~\ref{figure1} we present our most striking result. The figure
represents, in a color-coded scale, the average level of cooperation as a
function of the fraction of mostly cooperators, $\rho_C$, and mostly defectors,
$\rho_D$, for a BA network of contacts. The same plots but for the
rest of topologies explored (lattice and ER graphs) produce indistinguishable
results with respect to those shown in the figure. We therefore conclude
that the average level of cooperation in the system \emph{does not} depend
on the underlying structure. This means that, under the assumption that the
players follow the behavior of the experiment in \cite{grujic:2011b},
\emph{there is no network reciprocity}, i.e., no matter what the network of
contacts looks like, the observed level of cooperation is the same. This latter
finding is in stark contrast to most previous results coming out from numerical
simulations of models in which many different updating rules ---all of them based
upon the relative payoffs obtained by the players--- have been explored.

\begin{figure}
\begin{center}
\includegraphics[width=\columnwidth]{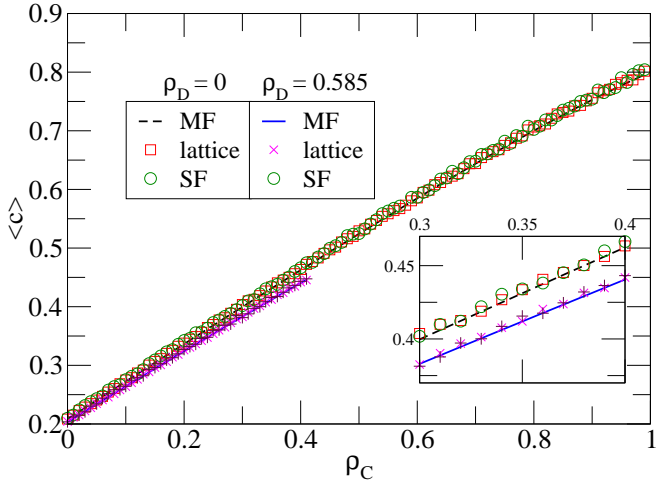}
\end{center}
\caption{(Color online) Average cooperation level in the stationary
state, $\langle c\rangle$, as a function of the density $\rho_C$
of mostly cooperators and two different values of the density $\rho_D$
of mostly defectors, for two different kinds of networks: regular
lattice ($k=8$), and Barab\'asi-Albert network ($\langle k\rangle=8$).
The network size is $N=10^4$ and the rest of parameters are as in
Fig.~\ref{figure1}. Lines represent the mean-field estimations. Results
are averages over 200 realizations. The inset is a zoom that highlights
how the different curves compare.}
\label{figure2}
\end{figure}

The previous numerical findings can be recovered using a simple
mean-field approach to the problem. Let the fractions of the three
types of players be $\rho_C$, $\rho_D$ and $\rho_X$, for mostly cooperators,
mostly defectors, and moody conditional cooperators, respectively, with the
obvious constraint $\rho_X=1-\rho_D-\rho_C$.  Denoting by $P_t(A)$ the
cooperation probability at time $t$ for strategy $A(=C,D,X)$
of the repeated PD we have
\begin{equation}
\langle c\rangle_t = \rho_CP(C)+\rho_DP(D)+\rho_XP_t(X),
\label{eq:avec}
\end{equation}
where $P_t(C)=P(C)$ and $P_t(D)=P(D)$ are known constants [in our case
$P(C)=0.8$, $P(D)=0.2$].
The probability of cooperation for conditional players in the next time
step can be obtained as
\begin{equation}
P_{t+1}(X)=(d+\gamma \langle c\rangle_t)P_t(X)+(a+\beta \langle c\rangle_t)
[1-P_t(X)],
\label{eq:PX}
\end{equation}
where the first term in the right hand side considers the probability that a
conditional cooperator keeps playing as a cooperator, whereas the second terms
stands for the situation in which a moody conditional cooperator switched from
defection to cooperation. Asymptotically
\[
\lim_{t\to\infty}P_t(X)=P(X), \qquad 
\lim_{t\to\infty}\langle c\rangle_t =\langle c\rangle.
\]
From Eq.~\eqref{eq:PX},
\begin{equation}
P(X)=\frac{a+\beta\langle c\rangle}{1+a-d+(\beta-\gamma)\langle c\rangle},
\end{equation}
thus \eqref{eq:avec} implies (with the replacement $\rho_X=1-\rho_C-\rho_D$)
\begin{equation}
A\rho_C+B\rho_D=1,
\label{SymbolicMeanFieldEquation}
\end{equation}
where
\begin{equation}
A\equiv\frac{P(C)-P(X)}{\langle c\rangle-P(X)}, \qquad
B\equiv\frac{P(D)-P(X)}{\langle c\rangle-P(X)},
\end{equation}
are functions of $\langle c\rangle$. From Eq.~\eqref{SymbolicMeanFieldEquation}
it follows that the curves of constant $\langle c\rangle$ are straight
lines in the simplex. Figure~\ref{figure1} clearly demonstrates this
fact: The straight lines are plots of Eq.~\eqref{SymbolicMeanFieldEquation}
for different values of $\langle c\rangle$. It can be seen that they are
parallel to the color stripes, and that the values of $\langle c\rangle$
they correspond to accurately fit those of the simulations.
Figure~\ref{figure2} depicts the curve $\langle c\rangle$ vs.~$\rho_C$
for two different values of $\rho_D$, as obtained from
Eq.~\eqref{SymbolicMeanFieldEquation} and compared to simulations.
This figure illustrates the excellent quantitative agreement between the
mean-field result and the simulation results.
The match between the analytical and numerical results is remarkable,
as it is the fact that the result does not depend on the 
underlying topology. This is the ultimate consequence of the lack
of network reciprocity: the cooperation level on any network can be
accurately modeled as if individuals were playing in a well-mixed
population. 

\begin{figure}
\begin{center}
\includegraphics[width=\columnwidth]{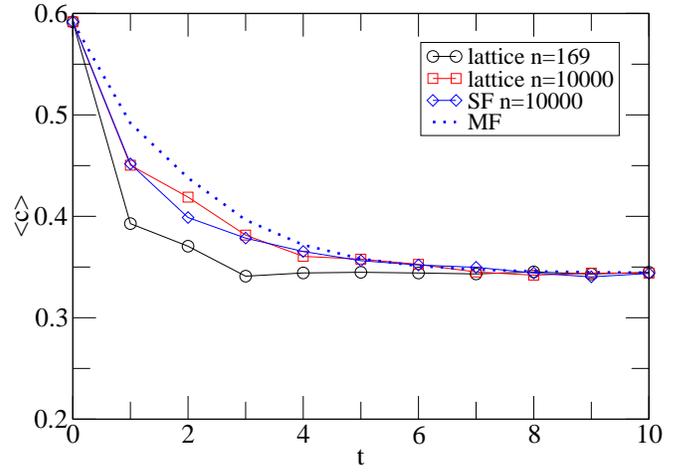}
\end{center}
\caption{(Color online) Time evolution of the cooperation level until the
stationary state is reached. The results have been obtained from numerical
simulations on different networks with different sizes. The Mean-Field
curve is the solution of Eq.~\eqref{eq:PX}. $P(C)=2/3$, $P(D)=1/3$,
$P(X;t=0)=1$, $\langle k\rangle=8$, $\rho_D=0.586$, $\rho_C=0.053$, $d=0.345$,
$a=0.224$, $\gamma=0.64$, $\beta=-0.072$. Averages have been taken over
$10^3$ realizations.}
\label{figure3}
\end{figure}

The steady state is reached after a rather short transient, as illustrated
by Figure~\ref{figure3}. This figure compares the approach of the cooperation
level to its stationary state as obtained iterating Eq.~\eqref{eq:PX} and
from numerical simulations on different networks with different sizes. The
initial cooperation level has been set to $\langle c\rangle_0=0.592$, close
to the value observed in the experiment of Ref.~\cite{grujic:2010}. The transient
does exhibit a weak dependence on the underlying topology and specially
on the network size, but for the largest simulated size ($N=10^4$) the
curves are all very close to the mean-field prediction.

\begin{figure}
\begin{center}
\includegraphics[width=0.9\columnwidth,clip=]{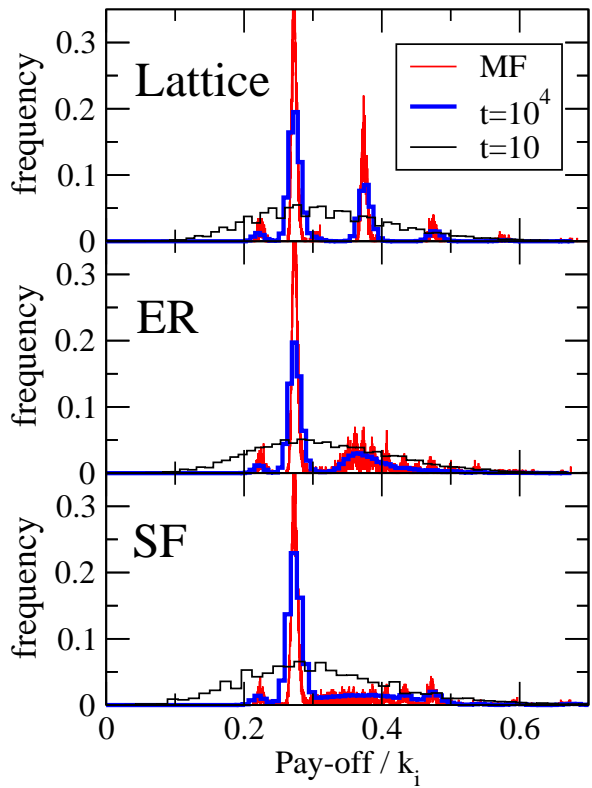}
\end{center}
\caption{(Color online) Distribution of the pay-off per
neighbor in the stationary state for different network topologies:
regular lattice ($k=8$), Erd\"{o}s R\'enyi ($\langle k\rangle=8$)
and Barab\'asi-Albert network ($\langle k\rangle=8$). Black and blue
lines represent the results of numerical simulations for two
values of time: $t=10$ (black shallow curves) and $t=10^4$ (blue, thick
line curves) while red lines represent the theoretical estimations for
the density probabilities at $t=10^4$, as obtained from
Eq.~\eqref{eq:Payoff:dis}. $N=10^4$, $\rho_D=0.586$,
$\rho_C=0.053$, and other parameters are as in Fig.~\ref{figure1}.
The simulation results are averages over $10^3$ realizations.}
\label{figure4}
\end{figure}

The only observable on which the topology does have a strong effect
is the payoff distribution among players. Figure~\ref{figure4} shows
these distributions for the three studied topologies, and at two different
times ---short and long. Smooth at short times, this distribution peaks
around certain values at long times. This reflects the fact that payoffs
depend on the number of neighbors of different types around a given
player, which yields a finite set of values for the payoffs (the centers
of the peaks). These numbers occur with different probabilities (determining
the height of the peaks), according to the distribution
\begin{equation}
Q(\mathbf{k})=\sum_{k\ge 1}\binom{k}{k_C\ k_D}\rho_C^{k_C}\rho_D^{k_D}
\rho_X^{k_X}p(k),
\label{eq:pd}
\end{equation}
where $p(k)$ is the degree distribution of the network and $\mathbf{k}=
(k_C,k_D,k_X)$, but it is understood that $k_X=k-k_C-k_D$. The
standard convention is assumed that the multinomial coefficient
$\binom{k}{k_C\ k_D}=0$ whenever $k_C<0$, $k_D<0$ or $k_X<0$.

The approach to a stationary distribution of payoffs exhibits a much
longer transient. This is due to the fluctuations in the payoffs arising
from the specific actions (cooperate or defect) taken by the players.
These fluctuations damp out as the accumulated payoffs approach their
asymptotic values. Thus, the peak widths shrink proportionally to $t^{-1/2}$. 
In fact, one can show that the probability density for
the distribution of payoffs $\Pi$ for strategy $Z$ can be approximated as
\begin{equation}
W_Z(\Pi)=\sum_{\mathbf{k}\ge 1}G\big(\Pi-a_k(Z)\mu(\mathbf{k}),
\sqrt{t}a_k(Z)\sigma(\mathbf{k})\big)Q(\mathbf{k}),
\label{eq:Payoff:dis}
\end{equation}
where $G(x,\gamma) \equiv (2\pi\gamma^2)^{-1/2}e^{-x^2/2\mathbf{\gamma}^2}$,
the mean payoff per neighbor received by a $Z$ strategist against a
cooperator is
\begin{equation*}
a_k(Z) \equiv \frac{1}{k}\{ P(Z)+T[1-P(Z)] \},
\end{equation*}
with $k=k_C+k_D+k_X$, and the average cooperation level in the neighborhood
of the focal player and its variance are
\begin{eqnarray*}
\mu(\mathbf{k}) & \equiv & k_CP(C)+k_DP(D)+k_XP(X), \\
\sigma(\mathbf{k})^2 & \equiv & k_C P(C)[1-P(C)]+k_D P(D)[1-P(D)]
\nonumber \\
& &+k_X P(X)[1-P(X)].
\end{eqnarray*}
The approximate total payoff distribution, $W(\Pi)=\rho_CW_C(\Pi)+
\rho_DW_D(\Pi)+\rho_XW_X(\Pi)$, is compared in Fig.~\ref{figure4}
with the results of the simulations for the longest time.

Summarizing, in this work we have shown both analytically and through
numerical simulations that if we take into account the way in which
humans are experimentally found to behave when facing social dilemmas
on lattices, no evidence of network reciprocity is obtained.
In particular, we have argued that if the players of a
Prisoners' Dilemma adopt an update rule that only depends on what they
see from their neighborhood, then cooperation drops to a low level
---albeit nonzero--- irrespective of
the underlying network. Moreover, we have shown
that the average level of cooperation obtained from
simulations is very well predicted by a mean-field model, and it is
found to depend only on the
fractions of different strategists.
Additionally, we have also shown that the underlying network of
contacts does manifest itself in the distribution of payoffs obtained
by the players, and has a slight influence on the transient behavior.

To conclude, it is worth mentioning that our results only make
sense when applied to evolutionary game models aimed at mimicking human
behavior in social dilemmas. The independence on the topology seems to
reflect the fact that humans update their actions according to a rule
that ignores relative payoffs. Interestingly, absence of network
reciprocity has also been observed in numerical simulations using best
response dynamics \cite{roca2009}, an update rule widely used in
economics that does not take into account the neighbors's payoffs. 
This suggests that the
result that networks do not play any role in the repeated PD may be
general for any dynamics that does not take neighbors' payoffs
into account. We
want to stress that the same kind of models thought of in a strict
biological context are ruled by
completely different mechanisms which do take into account payoff (fitness)
differences. Therefore, in such contexts lattice
reciprocity does play its role. In any case, our results call for
further experiments that uncover what rules are actually governing the behavior
of players engaged in this and other social dilemmas.

 
J.\ A.\ C.\ and A.\ S.\ acknowledge grants MOSAICO, PRODIEVO and Complexity-NET
RESINEE (Ministerio de Ciencia e Innovaci\'on, Spain) and MODELICO-CM
(Comunidad de Madrid, Spain). Y.\ M.\ was partially supported by Spanish MICINN
(Ministerio de Ciencia e Innovaci\'on) projects FIS2008-01240 and
FIS2009-13364-C02-01, by the FET-Open project
DYNANETS (grant no. 233847) funded by the European Commission and by
Comunidad de Arag\'on (Spain) through the project FMI22/10.


\end{document}